\documentclass[amsthm,seceqn,secthm,twoside]{elsart}

\usepackage{upref,amsmath,amssymb}

\usepackage{for-arXiv}

\makeatletter

\def\one{\@ifundefined{comp}{\kern.5pt\leavevmode\hbox{\upshape{\small1\kern-3.35pt\normalsize1}}}{\kern.5pt\mathbb{1}}}%
\def\l2zd{\ell^{2}(\mathbb{Z}^{d})}
\def\ltwoL{\ell^{2}(\Lambda)}
\def\zd{\mathbb{Z}^{d}}
\def\DN{\Delta_{\mathrm{N}}}
\def\DD{\Delta_{\mathrm{D}}}
\def\DX{\Delta_{\mathrm{X}}}
\def\DDT{\Delta_{\,\widetilde{\mathrm{D}}}}
\def\XinXset{\mathrm{X}\in\{\mathrm{N},\widetilde{\mathrm{D}},\mathrm{D}\}}
\def\NX{N_{\mathrm{X}}}
\def\NN{N_{\mathrm{N}}}
\def\NNinf{N_{\mathrm{N},\infty}}
\def\NNfin{N_{\mathrm{N, fin}}}
\def\ND{N_{\mathrm{D}}}
\def\NDT{N_{\widetilde{\mathrm{D}}}}

\def\e{\mathrm{e}}
\def\d{\mathrm{d}\hspace{1pt}}

\def\spec{\mathop\mathrm{spec}}
\def\dist{\mathop\mathrm{dist}}
\def\tr{\mathop\mathrm{trace}}

\def\DDTL{\Delta_{\,\widetilde{\mathrm{D}}, \Lambda}}

\def\LL{\mathbb{L}_{\Lambda}^{d}}
\def\EGL{\mathcal{E}_{\Lambda}}
\def\ELL{\mathbb{E}_{\Lambda}^{d}}
\def\HL{H_{\Lambda}}
\def\HG{\mathfrak{H}_{\mathfrak{G}}}
\def\hate{\mathfrak{h}(t)}

\def\le{\leqslant}
\def\ge{\geqslant}

\newcommand{\margintext}[1]          
   {\mbox{}\marginpar{\raggedright\hspace{0pt}\itshape\scriptsize#1}}
\newcounter{numcount}
\newcommand{\labelnummer}{\textup{(\roman{numcount})}}%
\newcommand{\indlabelnummer}{\textup{(\alph{numcount})}}%

\newenvironment{nummer}%
{\let\curlabelspeicher\@currentlabel%
  \begin{list}{\labelnummer}{\usecounter{numcount}\leftmargin0pt%
      \topsep0.5ex\partopsep2ex\parsep0pt\itemsep1ex\@plus1\p@%
      \labelwidth3.5em\itemindent4.5em\labelsep1em}%
    \let\saveitem\item%
    \def\item{\saveitem%
      \def\@currentlabel{\curlabelspeicher$\,$\labelnummer}%
      \let\label\bemlabel}}%
  {\end{list}}%

\AtBeginDocument{\let\mysaveref\ref}
\def\itemref#1{\mysaveref{#1item}}

\AtBeginDocument{\let\yetanotherlabel\label}

\def\bemlabel#1{\yetanotherlabel{#1}
  \def\@currentlabel{\labelnummer}
  \yetanotherlabel{#1item}}%

\newenvironment{indentnummer}%
    {\let\curlabelspeicher\@currentlabel%
      \begin{list}{\indlabelnummer}{\usecounter{numcount}%
                  \topsep1ex\partopsep2ex\parsep0pt\itemsep1ex
                  \labelwidth1.9em\itemindent0em\labelsep.6em%
                  \leftmargin2.5em}
      \let\saveitem\item%
      \def\item{\saveitem%
        \def\@currentlabel{\curlabelspeicher\indlabelnummer}%
        \let\label\indbemlabel}}%
    {\end{list}}%

\def\indbemlabel#1{\yetanotherlabel{#1}
  \def\@currentlabel{\indlabelnummer}
  \yetanotherlabel{#1item}}%

\newenvironment{remarks}{\begin{myremarks}\begin{nummer}}%
    {\end{nummer}\end{myremarks}}  

\theoremstyle{definition}
\newtheorem{myremarks}[thm]{Remarks}

\newif\ifper\pertrue
\def\per{.}

\def\au#1#2{{#1 #2}}
\def\lau#1#2{{#1 #2},}
\def\et{, }
\def\ti#1{#1,\ifper\fi\pertrue}

\def\bti{\@ifnextchar[\bbti\bbbti}
\def\bbti[#1]#2{{#2}, #1,}
\def\bbbti#1{{#1},}

\def\z{\@ifnextchar[\zz\zzz}
\def\zz[#1]#2#3#4#5{\perfalse{#2} {#3} (#5) #4 [#1]}
\def\zzz#1#2#3#4{{#1} {#2} (#4) #3\ifper\per\fi\pertrue}

\def\pub{\@ifstar\pubstar\pubnostar}
\def\pubnostar{\@ifnextchar[\@@pubnostar\@pubnostar}
\def\@@pubnostar[#1]#2#3#4{#2, #3, #4 #1\ifper\per\fi\pertrue}
\def\@pubnostar#1#2#3{#1, #2, #3\ifper\per\fi\pertrue}
\def\pubstar[#1]#2#3#4{\perfalse #2, #3, #4 [#1]\per\pertrue}

\makeatother

\begin{document}

\begin{frontmatter}

\title{Spectral asymptotics of the Laplacian on supercritical
    bond-percolation graphs\thanksref{support}}

\thanks[support]{Work supported by the Deutsche Forschungsgemeinschaft under
  grant nos.\  Mu~1056/2-1 and Sto~294/3-1.}

\author[pm]{Peter M\"uller}\ead{peter.mueller@physik.uni-goe.de}~and 
\author[ps]{Peter Stollmann}\ead{p.stollmann@mathematik.tu-chemnitz.de}
\address[pm]{Institut f\"ur Theoretische Physik, Georg-August-Universit\"at,
  37077 G\"ottingen, Germany}
\address[ps]{Technische Universit\"at Chemnitz, Fakult\"at f\"ur Mathematik,
  09107 Chemnitz, Germany}

\received{\today}



\begin{abstract}
  We investigate Laplacians on supercritical bond-percolation graphs with
  different boundary conditions at cluster borders. The integrated
  density of states of the Dirichlet Laplacian is found to exhibit a Lifshits
  tail at the lower spectral edge, while that of the Neumann Laplacian shows a
  van Hove asymptotics, which results from the percolating cluster. At the
  upper spectral edge, the behaviour is reversed.  
\end{abstract}

\end{frontmatter}

%
\section{Introduction and summary}
\label{intro}
%

Ever since Mark Kac posed the question ``Can one hear the shape of a drum?''
\cite{Kac66}, there has been a great deal of interest in finding relations
between the geometry of a manifold or a graph and spectral properties of the
Laplacian defined on it. The impressive works
\cite{Cou96,CoGr97,CoGr98,ChGr00,BaCo01}, which have been chosen by way of
example, witness the steady progress achieved in recent years and provide
further references.  Whereas Laplacians on manifolds dominated the scene in
the earlier years, the rise of spectral graph theory
\cite{Moh91,Mer94,CvDo95,Chu97,Col98} in the late 1980s and 90s has
contributed to deepen our understanding of the discrete case.

Spectral theory of random graphs, however, is still a widely open
field. The very recent contributions \cite{MaRe04,Bar04,HeHo05} take a
probabilistic point of view to derive heat-kernel estimates for Laplacians on
\emph{supercritical} Bernoulli bond-percolation graphs in the $d$-dimensional
hyper-cubic lattice. On the other hand, traditional methods from spectral
theory are used in \cite{KiMu04} to investigate the integrated density of
states of Laplacians on \emph{subcritical} bond-percolation graphs. Depending
on the boundary condition that is chosen at cluster borders, two different
types of \emph{Lifshits asymptotics} at spectral edges were found
\cite{KiMu04}. For example, the integrated density of states of the Neumann
Laplacian behaves as
\begin{equation}
  \label{NNsub}
  \text{``~}\NN(E) -\NN(0) \sim \exp\{ -E^{-1/2}\}\text{~''} \qquad\text{as~~}
  E\downarrow 0 
\end{equation}
at the lower spectral edge for bond probabilities $p$ below the percolation
threshold $p_{c}$. We have put quotation marks here, because,
strictly speaking, one should take appropriate
logarithms on both sides. The Lifshits exponent $1/2$ in
\eqref{NNsub} is independent of the spatial dimension $d$. This was explained
by the fact that, asymptotically, $\NN$ is dominated by the smallest
eigenvalues which arise from very long \emph{linear clusters} in this case.
In contrast, for the Dirichlet Laplacian and $p < p_{c}$, it was found that
\begin{equation}
  \label{NDsub}
  \text{``~}\ND(E)  \sim \exp\{ -E^{-d/2}\}\text{~''} \qquad\text{as~~}
  E\downarrow 0 \,.
\end{equation}
We note that $\ND(0)=0$. The Lifshits exponent in \eqref{NDsub} comes out as
$d/2$, because the dominating small Dirichlet eigenvalues arise from large
\emph{fully connected cube- or sphere-like clusters}. Thus, depending on the
boundary condition (and the spectral edge) different geometric graph
properties show up in the integrated density of states. We refer to the
literature cited in \cite{KiMu04} for a discussion of other spectral
properties of these and closely related operators, for the history of the
problem and what is known in the physics literature. Lifshits asymptotics for
a Neumann Laplacian on Erd\H{o}s--R\'enyi random graphs are studied in
\cite{KhKi05}.

In this paper we pursue the investigations of \cite{KiMu04} and ask what
happens to \eqref{NNsub} and \eqref{NDsub} in the \emph{supercritical phase}
of bond-percolation graphs. Clearly, one would not expect the contribution of
the finite clusters to alter the picture completely. But for the infinite
percolating cluster, the story may be different. Indeed, we will prove that
the percolating cluster produces a \emph{van Hove asymptotics}
\begin{equation}
  \label{NNsup}
  \text{``~}\NN(E) -\NN(0) \sim E^{d/2}\text{~''} \qquad\text{as~~}
  E\downarrow 0 
\end{equation}
in the Neumann case for $p>p_{c}$. There is also an additional Lifshits-tail
behaviour due to finite clusters, but it is hidden under the dominating
asymptotics \eqref{NNsup}. Loosely speaking, \eqref{NNsup} is true because the
percolating cluster looks like the full regular lattice on very large length
scales (bigger than the correlation length) for $p>p_{c}$. On smaller scales
its structure is more like that of a jagged fractal. The Neumann Laplacian
does not care about these small-scale holes, however. All that is needed for
\eqref{NNsup} to be true is the existence of a suitable $d$-dimensional,
infinite grid.  In contrast, the Dirichlet Laplacian does care about holes at
all scales so that \eqref{NDsub} continues to hold for $p \ge p_{c}$, as we
shall prove.  Low-lying Dirichlet eigenvalues require large fully connected
cube- or sphere-like regions, and this is a large-deviation event.

Closely related large-deviation results for Laplacians on percolation
graphs have been obtained in \cite{Ant95,BiKo01}. To be precise,
\cite{Ant95,BiKo01} refer to the Pseudo-Dirichlet Laplacian $\DDT$ in
the sense of our Definition~\ref{deltadef}(ii) below. Considering both
site- and bond-percolation graphs, and using a discrete version of the
method of enlargement of obstacles, Antal \cite{Ant95} derives the
long-time asymptotics for the mean (i.e.\ annealed) hitting-time
distribution of the set of absent sites (resp.\ bonds) for the random
walk generated by $\DDT$. Biskup and K\"onig work in the setting of
the parabolic Anderson model, which contains $\DDT$ on
\emph{site}-percolation graphs as a special case. In particular, they
establish a Lifshits tail for the corresponding integrated density of states,
see also Remark~\ref{bikorem}.

This paper is organised as follows. In the next section we give a precise
statement of our results in Theorems~\ref{lifshits} and~\ref{hove}.
Section~\ref{lifshitsproof} is devoted to the proof of Theorem~\ref{lifshits}.
In this proof we follow the strategy laid down in \cite{Sto99}, see also
\cite{Sto01}. The goal there was to establish Lifshits tails in the context
of random Schr\"odinger operators.  Finally, Section~\ref{hoveproof} contains
the proof of Theorem~\ref{hove}, where we apply the recent deep heat-kernel
estimates from \cite{MaRe04,Bar04,HeHo05}.

%
\section{Definitions and precise formulations} \label{defres}
%

To set up the mathematical arena, let us first recall some notions from
Bernoulli bond percolation.  For $d \in\mathbb{N\,}$, a natural number, we
denote by $\mathbb{L}^{d}$ the (simple hyper-cubic) lattice in $d$ dimensions.
Being a graph, the lattice $\mathbb{L}^{d} = (\zd, \mathbb{E}^{d})$ has the
\emph{vertex set} $\zd$ and the \emph{edge set} $\mathbb{E}^{d}$ given by all
unordered pairs $\{x,y\}$ of nearest-neighbour vertices $x,y \in \zd$, that is,
those vertices which have Euclidean distance $|x-y| : = \bigl(\sum_{\nu=1}^{d}
|x_{\nu} -y_{\nu}|^{2}\bigr)^{1/2} = 1$. Here, elements of $\zd$ are
canonically represented as $d$-tuples $x=(x_{1},\ldots, x_{d})$ with entries
from $\mathbb{Z}$.  Next, we consider the probability space $\Omega
=\{0,1\}^{\mathbb{E}^{d}}$, which is endowed with the usual product
sigma-algebra, generated by finite cylinder sets, and equipped with a product
probability measure $\mathbb{P}$.  Elementary events in $\Omega$ are sequences
of the form $\omega \equiv (\omega_{\{x,y\}})_{\{x,y\} \in \mathbb{E}^{d}}$, and
we assume their entries to be independently and identically distributed
according to a Bernoulli law $\mathbb{P} (\omega_{\{x,y\}} =1 ) =p$ with
\emph{bond probability} $p \in ]0,1[$. To a given $\omega\in\Omega$, we
associate an edge set $\mathcal{E}^{(\omega)} := \bigl\{ \{x,y\} \in
\mathbb{E}^{d} : \omega_{\{x,y\}} =1 \bigr\}$.

A \emph{bond-percolation graph in
  $\zd$} is the mapping $\mathcal{G}: \Omega \ni \omega \mapsto
\mathcal{G}^{(\omega)} := (\mathbb{Z}^{d}, \mathcal{E}^{(\omega)})$ with values
in the set of subgraphs of $\mathbb{L}^{d}$. Given $x\in\zd$, the
\emph{vertex degree} $d_{\mathcal{G}^{(\omega)}}(x)$ counts the number of
edges in $\mathcal{G}^{(\omega)}$ which share $x$ as a common vertex. 
\smallskip

\begin{defn} \label{deltadef}
  The random operators $D:\Omega \ni \omega \mapsto D^{(\omega)}$,
  respectively $A:\Omega \ni \omega \mapsto A^{(\omega)}$, are called 
  vertex-degree operator, respectively adjacency operator, of bond-percolation
  graphs in $\zd$. Their realisations, $D^{(\omega)} : \l2zd \rightarrow
  \l2zd$, respectively $A^{(\omega)} : \l2zd \rightarrow \l2zd$, act on the
  Hilbert space of complex-valued, square-summable sequences indexed by $\zd$
  according to 
  \begin{equation}
    \begin{split}
      D^{(\omega)}\varphi (x) &:= d_{\mathcal{G}^{(\omega)}}(x) \,\varphi(x)\,,
      \\
      A^{(\omega)}\varphi(x)  &:= \sum_{y \in \zd:\; \{x,y\} \in
        \mathcal{E}^{(\omega)}}\varphi(y)\,,
    \end{split}
  \end{equation}
  for all $\varphi\in\l2zd$, all $x\in\zd$ and all $\omega\in\Omega$. With
  these definitions, we introduce \emph{Laplacians on bond-percolation graphs}
  for three different ``boundary conditions'' at
  non-fully connected vertices \\[.5ex]
  \begin{tabular}{@{}r@{\quad}l@{\quad}l@{}}
    \textup{(i)}  & \emph{Neumann Laplacian}: 
                          & $\DN := D - A,$\\ 
    \textup{(ii)} & \emph{Pseudo-Dirichlet Laplacian}: 
                          & $\DDT := \DN + (2d \one - D) =  2d \one  - A,$\\
    \textup{(iii)} & \emph{Dirichlet Laplacian}: 
                          & $\DD := \DN + 2 (2d \one - D).$
  \end{tabular}
  \par\noindent 
  Here $\one$ stands for the identity operator on $\l2zd$.
\end{defn}

\begin{remarks}
\item The motivation and origin of the terminology for the different 
  boundary conditions are discussed in \cite{KiMu04} -- together with some
  basic properties of the operators.
\item \label{genergodic} The random self-adjoint Laplacians are ergodic with
  respect to $\zd$-translations. Hence, their spectra and the spectral subsets
  arising in the Lebesgue decomposition are all equal to non-random sets with
  probability one. In particular, the spectrum is $\mathbb{P}$-almost surely
  given by $\spec (\DX) = [0,4d]$ for all $\XinXset$, as was shown in
  \cite{KiMu04}.
\end{remarks}

Next, we define the quantity of our main interest for this paper, the
integrated density of states of $\DX$. To this end let $\delta_{x} \in \l2zd$
be the sequence which is concentrated at the point $x\in\zd$, i.e.\ 
$\delta_{x}(x) :=1$ and $\delta_{x}(y) := 0$ for all $y \in\zd \setminus\{x\}$.
Moreover, $\Theta$ stands for the Heaviside unit-step function, which we
choose to be right continuous, viz.\ $\Theta(E) := 0$ for all real $E<0$ and
$\Theta (E) := 1$ for all real $E \ge0$.
\smallskip

\begin{defn}
  \label{Ndef}
  For every $p \in ]0,1[$ and every $\XinXset$ we call the function 
  \begin{equation}
    \NX : \mathbb{R} \ni E \mapsto \NX(E) :=
    \int_{\Omega}\!\mathbb{P}(\d\omega)  \; \langle\delta_{0} ,  
    \Theta\bigl(E - \DX^{(\omega)}\bigr) \delta_{0}\rangle 
  \end{equation}
  with values in the interval
  $[0,1]$ the \emph{integrated density of states of} $\DX$.
\end{defn}

\begin{remarks}
\item The integrated density of states $\NX$ is the right-conti\-nuous
  distribution function of a probability measure on $\mathbb{R}$. The set of
  its growth points coincides with the $\mathbb{P}$-almost-sure spectrum
  $[0,4d]$ of $\DX$.
\item 
  It is shown in \cite{KiMu04} that the Laplacians are related to each
  other by a unitary involution, which implies the symmetries
  \begin{equation}
    \label{bcrel}
    \begin{split}
      \NDT (E) &= 1 - \lim_{\varepsilon\uparrow 4d -E} \NDT(\varepsilon)\,,
      \\
      N_{\mathrm{D(N)}}(E) &= 1 - \lim_{\varepsilon\uparrow 4d -E}
      N_{\mathrm{N(D)}} (\varepsilon)
    \end{split}
  \end{equation}
  for their integrated densities of states for all $E\in\mathbb{R}$. The
  limits on the right-hand sides of \eqref{bcrel} ensure that the discontinuity
  points of $\NX$ are approached from the correct side.
\item \label{fininf}
  By ergodicity, Definition~\ref{Ndef} of the integrated density of states
  coincides with the usual one in terms of a macroscopic limit of a
  finite-volume eigenvalue counting function. More precisely, let $\Lambda
  \subset\zd$ stand for bounded cubes centred at the origin with volume
  $|\Lambda|$. For every $\XinXset$ let $\Delta_{\mathrm{X},\Lambda}$ be the
  finite-volume restriction of $\DX$ to $\ell^{2}(\Lambda)$ introduced in
  Def.~{\normalfont 1.11} in \cite{KiMu04}.  Then there exists a set $\Omega' \subset
  \Omega$ of full probability, $\mathbb{P}(\Omega') =1$, such that
  \begin{equation}
    \label{Nlimit}
    \NX(E) = \lim_{\Lambda \uparrow\zd} \left[\frac{1}{|\Lambda|} \;
    \tr\nolimits^{\phantom{y}}_{\ell^{2}(\Lambda)} \Theta \bigl(E-
    \Delta^{(\omega)}_{\mathrm{X}, \Lambda}\bigr) \right]
  \end{equation}  
  holds for all $\omega \in \Omega'$ and all $E\in\mathbb{R}\,$, except for
  the (at most countably many) discontinuity points of $\NX$, see
  Lemma~{\normalfont 1.12}
  in \cite{KiMu04}. In Section~\ref{lifshitsproof} we will construct another
  finite-volume restriction of $\DDT$, for which \eqref{Nlimit} holds, too.
\end{remarks}

Let $p_{c}\equiv p_{c}(d)$ denote the critical bond probability of the
percolation transition in $\zd$. We recall that $p_{c}=1$ for $d=1$, otherwise
$p_{c} \in]0,1[$, see e.g.\ \cite{Gri99}.  Despite the title of this paper,
our first main result covers the non-percolating phase $p\in]0,p_{c}[$ and the
critical point $p=p_{c}$, too.
\smallskip

\begin{thm}
  \label{lifshits}
  Assume $d\in\mathbb{N}$ and $p \in ]0,1[$. Then the integrated
  density of states $\NX$ of the Laplacian $\DX$ on bond-percolation graphs
  in $\zd$ exhibits a \emph{Lifshits tail} at the lower
  spectral edge 
  \begin{equation}
    \label{lowerlif}
    \lim_{E\downarrow 0}\;\frac{\ln | \ln \NX(E)|}{\ln E} =
    -\; \frac{d}{2}  \qquad \mathit{for} \quad
    \mathrm{X} \in \{\widetilde{\mathrm{D}}, \mathrm{D}\}
  \end{equation}
  and at the upper spectral edge
  \begin{equation}
    \label{upperlif}
    \lim_{E \uparrow 4d}\;\frac{\ln |\ln [1 - \NX(E)]|}{\ln (4d -E)} =  
    -\; \frac{d}{2}  \qquad \mathit{for} \quad
    \mathrm{X} \in \{\mathrm{N}, \widetilde{\mathrm{D}}\} \,.
  \end{equation}
\end{thm}

\begin{remarks}
\item 
  The theorem follows directly from the upper and lower bounds in
  Lemma~\ref{liflemma} below, together with the subsequent
  Remark~\ref{lifrelate}. In fact, the bounds of Lemma~\ref{liflemma} provide
  a slightly stronger statement than Theorem~\ref{lifshits}.
\item \label{lifrelate}
  The Lifshits tails at the upper spectral edge are related to the ones at
  the lower spectral edge by the symmetries \eqref{bcrel}.
\item In the non-percolating phase, $p\in]0,p_{c}[$, the content of the
  theorem is known from \cite{KiMu04}, where it is proved by a different
  method. The method of \cite{KiMu04}, however, does not seem to extend to the
  critical point or the percolating phase, $p \in]p_{c},1[$.
\item 
  The Lifshits asymptotics of Theorem~\ref{lifshits} are determined by those
  parts of the percolation graphs, which contain large, fully-connected cubes.
  This also explains why the spatial dimension enters the Lifshits exponent
  $d/2$.
\item \label{bikorem}
  We expect that \eqref{lowerlif} can be refined in the case $\mathrm{X} =
  \widetilde{\mathrm{D}}$ as to obtain the constant 
  \begin{equation}
    \lim_{E\downarrow 0}\; \frac{\ln \NDT(E)}{E^{-d/2}}=: -c_{*}(d,p)\,.
  \end{equation}
  An analogous statement is known from Thm.~{\normalfont 1.3} in
  \cite{BiKo01} for the case of \emph{site}-percolation graphs.
  Moreover, it is demonstrated in \cite{Ant95} that the bond- and the
  site-percolation cases have similar large-deviation properties.
\end{remarks}

Our second main result complements Theorem~\ref{lifshits} in the percolating
phase. 
\smallskip

\begin{thm}
  \label{hove}
  Assume $d \in\mathbb{N}\setminus\{1\}$ and $p \in ]p_{c},1[$. Then the
  integrated density of states of the Neumann Laplacian $\DN$ on
  bond-percolation graphs in $\zd$ exhibits a \emph{van Hove asymptotics} at
  the lower spectral edge
  \begin{equation}
    \label{lowerhove}
    \lim_{E\downarrow 0}\;\frac{\ln[ \NN(E) - \NN(0)]}{\ln E} =
    \frac{d}{2} \,,
  \end{equation}
  while that of the Dirichlet Laplacian $\DD$ exhibits one at the upper
  spectral edge
  \begin{equation}
    \label{upperhove}
    \lim_{E \uparrow 4d}\;\frac{\ln [\ND^{-}(4d) - \ND(E)]}{\ln (4d -E)} =  
    \frac{d}{2} \,,
  \end{equation}
  where $\ND^{-}(4d) := \lim_{E\uparrow 4d} \ND(E) = 1- \NN(0) $. 
\end{thm}

\begin{remarks}
\item The theorem follows directly from the upper and lower bounds in
  Lemma~\ref{hovelemma} below, together with the symmetries \eqref{bcrel}.  In
  fact, the bounds of Lemma~\ref{hovelemma} provide a slightly stronger
  statement than Theorem~\ref{hove}.  Lemma~\ref{hovelemma} relies mainly on
  recent estimates \cite{MaRe04,Bar04,HeHo05} for the long-time decay of the
  heat kernel of $\DN$ on the percolating cluster.
\item 
  The reference value $\NN(0)$ in \eqref{lowerhove} results from the mean
  number density of zero eigenvalues of the Neumann Laplacian \cite{KiMu04}.
  It is given by
  \begin{equation}
    \NN(0) = \kappa(p) + (1-p)^{2d}\,,
  \end{equation}
  where $\kappa(p)$ is the mean number density
  of clusters, see e.g.\ Chap.~{\normalfont 4} in \cite{Gri99}, and
  $(1-p)^{2d}$ the mean number density of isolated vertices.
\item The counterpart of Theorem~\ref{hove} for the non-percolating phase,
  $p\in]0,p_{c}[$, was proved in \cite{KiMu04}. There, $\NN$ was shown to have
  a different kind of Lifshits asymptotics with a Lifshits exponent $1/2$ at
  the lower spectral edge, see also Section~\ref{intro}, and the same is true
  for $\ND$ at the upper spectral edge. This type of Lifshits behaviour is
  caused by large (isolated) linear clusters, which explains why the spatial
  dimension does not influence the Lifshits exponent. This behaviour is also
  present for $p\in]p_{c},1[$, but hidden under the more dominant van Hove
  asymptotics caused by the percolating cluster.
\item At the critical point $p=p_{c}$, the behaviour of $\NN$ at the lower
  spectral edge, respectively that of $\ND$ at the upper spectral edge, is an
  open problem.
\end{remarks}


\section{Proof of Theorem~\ref{lifshits}}
\label{lifshitsproof}

In this section we prove the Lifshits-tail behaviour of
Theorem~\ref{lifshits}. Thanks to the symmetries \eqref{bcrel}, it suffices to
consider the lower spectral edge only.
\smallskip

\begin{lem}
  \label{liflemma}
  For every $d\in\mathbb{N}$ and every $p\in]0,1[$ there exist constants
  $\varepsilon_{\mathrm{D}}$, $\alpha_{u}$, $\alpha_{l} \in ]0,\infty[$ such
  that 
  \begin{equation}
    \label{lifbounds}
    \exp\{ -\alpha_{l} E^{-d/2}\} \le \ND(E) \le \NDT(E) \le  \exp\{
    -\alpha_{u} E^{-d/2}\} 
  \end{equation}
  holds for all $E\in ]0, \varepsilon_{\mathrm{D}}[$.
\end{lem}

\begin{proof}
  The left inequality in \eqref{lifbounds}, i.e.\ the lower bound on $\ND$,
  was proved in Lemma~{\normalfont 2.9} in \cite{KiMu04}. The middle one simply
  reflects the operator inequality $\DDT^{(\omega)} \le \DD^{(\omega)}$, which
  is valid for all $\omega\in\Omega$. So it remains to prove the upper bound
  on $\NDT$. 
  
  We follow the strategy of the proof in \cite{Sto99}, see also
  Chap.~{\normalfont 2.1} in
  \cite{Sto01}. To do so, we have to fix some notation, first. Given a bounded
  cube $\Lambda \subset \zd$ and $x\in\Lambda$, we introduce the boundary
  degree
  \begin{equation}
    b_{\partial\Lambda}(x) := \bigl|\bigl\{ \{x,y\} \in \mathbb{E}^{d} : y \notin
    \Lambda \bigr\}\bigr| 
  \end{equation}
  as the cardinality of the set of edges in the regular lattice
  $\mathbb{L}^{d}$ that connect $x$ with $\zd\setminus \Lambda$. The
  restriction $\mathcal{G}_{\Lambda}^{(\omega)} := (\Lambda,
  \mathcal{E}_{\Lambda}^{(\omega)})$ with
  $\mathcal{E}_{\Lambda}^{(\omega)} := \bigl\{ \{x,y\}
  \in\mathcal{E}^{(\omega)} : x,y \in\Lambda\bigr\}$ of any
  realisation $\mathcal{G}^{(\omega)}$ of a bond-percolation graph to
  $\Lambda$ is obtained by keeping only vertices and edges within
  $\Lambda$, and $d_{\mathcal{G}_{\Lambda}^{(\omega)}} (x)\le 2d -
  b_{\partial\Lambda}(x)$ stands for the associated vertex degree of
  $x\in\Lambda$. In particular, $\mathbb{E}_{\Lambda}^{d} := \bigl\{
  \{x,y\} \in\mathbb{E}^{d} : x,y \in\Lambda\bigr\}$ is the edge set
  of the fully connected cube $\mathbb{L}_{\Lambda}^{d} := (\Lambda,
  \mathbb{E}_{\Lambda}^{d})$, that is the restriction of the regular
  lattice $\mathbb{L}^{d}$ to $\Lambda$.  Finally, let
  $\ell^{2}(\Lambda)$ be the Hilbert space of complex-valued (finite)
  sequences indexed by $\Lambda$, and, given any subgraph
  $\mathfrak{G} := (\Lambda,\mathfrak{E})$ of $\LL$, we introduce the
  operator $\mathfrak{H}_{\mathfrak{G}} : \ltwoL \rightarrow \ltwoL$,
  $\varphi\mapsto \HG\varphi$, where
  \begin{align}
    \label{hlambda}
    \HG\varphi (x) &:= 
    -  \sum_{y\in\Lambda : \{x,y\} \in \mathfrak{E}} \varphi(y) \; +
    \bigl( 2d - b_{\partial\Lambda}(x) \bigr) \varphi(x)
    \nonumber\\ 
    &\phantom{:}= \sum_{y\in\Lambda : \{x,y\} \in
      \mathfrak{E}} \bigl( \varphi(x) -\varphi(y)\bigr) \;
     + \bigl( 2d - b_{\partial\Lambda}(x) -
      d_{\mathfrak{G}} (x)\bigr) \varphi(x)
  \end{align}
  for all $\varphi\in\ell^{2}(\Lambda)$ and all $x\in\Lambda$. Now, we define
  the restriction of the Pseudo-Dirichlet Laplacian $\DDT$ to the cube
  $\Lambda$ with Neumann conditions along the boundary $\partial\Lambda$ of
  $\Lambda$ as the random bounded self-adjoint operator $H_{\Lambda}$ with
  realisations $H_{\Lambda}^{(\omega)} :=
  \mathfrak{H}_{\mathcal{G}_{\Lambda}^{(\omega)}}$ for all $\omega \in
  \Omega$. 
  
  Next we claim that 
  \begin{equation}
    \label{finiterep}
    \NDT(E) = \inf_{\Lambda \subset\zd} \left[\frac{1}{|\Lambda|} 
    \int_{\Omega}\!  \mathbb{P}(\d\omega') \;
    \tr\nolimits^{\phantom{y}}_{\ell^{2}(\Lambda)} \Theta (E-
    \HL^{(\omega')} ) \right]
  \end{equation}
  holds for all $E\in\mathbb{R}$. This is so, because (i)~~ the operator
  $\HL^{(\omega)}$ differs from the finite-volume restriction
  $\DDTL^{(\omega)}$ in Remark~\ref{fininf} by a perturbation whose rank is at
  most of the order of $|\partial\Lambda|$, the surface area of the cube
  $\Lambda$. Hence, \eqref{Nlimit} remains true for $\mathrm{X}
  =\widetilde{\mathrm{D}}$ and with $\DDTL^{(\omega)}$ being replaced by
  $\HL^{(\omega)}$ on its right-hand side,
  \begin{equation}
    \label{doubleerg}
    \NDT(E) = \lim_{\Lambda \uparrow\zd} \left[\frac{1}{|\Lambda|} \;
    \tr\nolimits^{\phantom{y}}_{\ell^{2}(\Lambda)} \Theta (E-
    \HL^{(\omega)} ) \right]\,.
  \end{equation}  
  (ii)~~ On the other hand, $\HL^{(\omega)}$ is designed in such a way that
  $H_{\Lambda_{1}}^{(\omega)} \oplus H_{\Lambda_{2}}^{(\omega)} \le
  H_{\Lambda_{1}\cup \Lambda_{2}}^{(\omega)}$ holds on
  $\ell^{2}(\Lambda_{1}\cup\Lambda_{2})$ for all bounded cubes
  $\Lambda_{1},\Lambda_{2} \subset \mathbb{Z}^{d}$ with $\Lambda_{1}\cap
  \Lambda_{2} = \varnothing$ and for all $\omega\in\Omega$. Hence, $\Theta (E-
  H_{\Lambda}^{(\omega)})$ gives rise to a subergodic process and we conclude
  from the Ackoglu--Krengel subergodic theorem that the right-hand side of
  \eqref{doubleerg} equals the right-hand side of \eqref{finiterep} -- again
  for all continuity points of the limit and uniformly for $\omega$ in a set
  of probability one.  (iii)~~ From this we have \eqref{finiterep} for all
  continuity points of both sides. But since both sides of \eqref{finiterep}
  are right-continuous functions of $E$, equality holds for all
  $E\in\mathbb{R}$, and the derivation is complete.

  From \eqref{finiterep} we infer the upper bound
  \begin{equation}
    \label{upperstart}
    \NDT(E) \le \inf_{\Lambda \subset\zd} \mathbb{P}[E_{\Lambda} \le E]\,,
  \end{equation}
  where the non-negative random variable $E_{\Lambda}$ stands for the smallest
  eigenvalue of the random operator $H_{\Lambda}$. 
  
  The aim is to obtain a simple large-deviation estimate for the probability
  in \eqref{upperstart}. This will be achieved with the help of analytic
  perturbation theory along the lines of \cite{Sto99}, see also Sec.\
  {\normalfont 4.1.10}
  in \cite{Sto01}. We write $H_{0,\Lambda} := \mathfrak{H}_{\LL}$ for the
  Neumann Laplacian of the fully connected cube $\LL$ and $W_{\Lambda} :=
  H_{\Lambda} - H_{0,\Lambda}$. Given $t\in [0,1]$, we 
  introduce 
  \begin{equation}
    \label{hael}
    \HL(t) := H_{0,\Lambda} + t W_{\Lambda}
  \end{equation}
  so that $\HL(1) = \HL$. We want to construct an upper bound for the
  probability that $E_{\Lambda}$ is small. Denoting the bottom eigenvalue of
  $\HL(t)$ by $E_{\Lambda}(t)$, we use the following ideas.
  \begin{indentnummer}
  \item \label{monotone}
    The function $[0,1] \ni t \mapsto E_{\Lambda}(t)$ is non-decreasing,
    $E_{\Lambda}(0) =0$ and $E_{\Lambda}(1)=E_{\Lambda}$.
  \item \label{aptapply}
    This function can be linearised, if its argument is small
    enough. More precisely, there exist constants $\tau, \beta \in
    ]0,\infty[$, which depend only on the spatial dimension $d$, such that 
    \begin{equation}
      \label{aptresult}
      | E_{\Lambda}(t) - t E'_{\Lambda}(0) | \le \beta t^{2} |\Lambda|^{2/d}
    \end{equation}
    for all $t \in [0, \tau|\Lambda|^{-2/d}]$. Here, we have set
    $E'_{\Lambda}(0) := \frac{\d}{\d t} E_{\Lambda}(t)\big|_{t=0}$.
  \item \label{largedev}
    The slope $E'_{\Lambda}(0)$ obeys a large-deviation
    estimate. Given any $\alpha \in ]0, 1-p[$, there exists a constant $\gamma
    \in ]0,\infty[$, which depends on $p$ and $d$, such that  
    \begin{equation}
      \label{ldresult}
      \mathbb{P} [ E'_{\Lambda}(0) \le \alpha ] \le \e^{-\gamma |\Lambda|}
      \,. 
    \end{equation}
  \end{indentnummer}
  We will prove \itemref{monotone} with a Perron--Frobenius argument in
  Lemma~\ref{groundstate} and discuss observations \itemref{aptapply} and
  \itemref{largedev} below. For the time being, let us go on to estimate the
  probability that $E_{\Lambda}$ is small. 

  Suppose $E_{\Lambda}(t) \le E$. Then we conclude from
  \itemref{monotone}, the triangle inequality and \itemref{aptapply}
  that
  \begin{equation}
    \label{aptstart}
    E_{\Lambda}'(0) \le  \frac{E_{\Lambda}(t)}{t}  + \Bigl|
      \frac{E_{\Lambda}(t)}{t} -   E_{\Lambda}'(0) \Bigr|   \le
    \frac{E}{t} + \beta t |\Lambda|^{2/d}\, , 
  \end{equation}
  provided $t$ is small enough. So we need to adjust $t\equiv t_{E}$ and
  $\Lambda \equiv\Lambda_{E}$ such that $t_{E} \le \tau |\Lambda_{E}|^{-2/d}$.
  Moreover, we aim to achieve that the right-hand side of \eqref{aptstart} is
  bounded from above by some $\alpha< 1-p$. This is accomplished in the
  following way. Without restriction we can assume that, in addition, $\alpha
  < 2\beta\tau$. Then we set $t_{E} := \alpha/(2\beta |\Lambda_{E}|^{2/d})$
  and choose the size of the cube such that
  \begin{equation}
    \label{lchoose}
    \frac{\alpha}{2 (\beta E)^{1/2}} -1 \le |\Lambda_{E}|^{1/d} \le 
    \frac{\alpha}{2 (\beta E)^{1/2}}\,.
  \end{equation}
  For this to make sense, the right-hand side of \eqref{lchoose} has to exceed
  2. So, we restrict ourselves to low energies, say $E \in ]0,
  \varepsilon_{\mathrm{D}}[$, and summarise this argument as 
  \begin{equation}
    \label{aptsumm}
    E_{\Lambda_{E}}(t_{E}) \le E \qquad\text{implies}\qquad 
    E_{\Lambda_{E}}'(0) \le \alpha < 1-p\,.
  \end{equation}
  Note that $\varepsilon_{\mathrm{D}}$ depends only on $p$ and $d$.

  Altogether, we infer from
  Eq.\ \eqref{upperstart}, observation~\itemref{monotone},
  implication \eqref{aptsumm} and observation~\itemref{largedev} that
  \begin{equation}
    \label{upperchain}
    \NDT(E) \le \mathbb{P}[ E_{\Lambda_{E}}(t_{E}) \le E ] \le 
    \mathbb{P}[ E'_{\Lambda_{E}}(0) \le \alpha ] \le
    \e^{-\gamma |\Lambda_{E}|} \le \e^{-\alpha_{u}E^{-d/2}} \,,
  \end{equation}
  where $\alpha_{u}\in ]0,\infty[$ is a constant that depends only on $p$ and
  $d$. 
  
  Next, we verify observations~\itemref{aptapply}
  and~\itemref{largedev} above. Observation~\itemref{aptapply} relies
  on a deterministic result from analytic perturbation theory. To this
  end we consider the operator family $H(z) := H_{0} + z H_{1}$ for
  $z\in\mathbb{C}$. Here, $H_{0} := H_{0,\Lambda}$ is the Neumann
  Laplacian of $\LL$ and $H_{1}:= W_{\Lambda}^{(\omega)}$ the
  perturbation with $\omega\in\Omega$ arbitrary, but fixed.  The
  bottom eigenvalue $0$ of $H_{0}$ is an isolated simple eigenvalue.
  Its isolation distance $\vartheta := \dist \bigl(0,
  \spec(H_{0}\setminus \{0\}) \bigr)$ is determined by the magnitude
  of the smallest non-zero eigenvalue of $H_{0}$. This distance obeys
  the estimate $\vartheta \ge c |\Lambda|^{-2/d}$ for some constant $c
  \in ]0,\infty[$, which follows from reducing the eigenvalue problem
  for $H_{0}$ to that of a linear chain by separation of variables and
  applying a Cheeger-type inequality, see e.g.\ {\normalfont(2.6)} in
  \cite{KiMu04}. Moreover, we have the uniform bound $\|
  W_{\Lambda}^{(\omega)}\| \le 2d$ for the operator norm of the
  perturbation so that $H(z)$ has one isolated eigenvalue $E(z)$ in
  the complex disc $B_{\vartheta /2}(0)$ provided $z < \vartheta
  /(4d)$. We refer to \cite{Kat76}, Chap.~II, \S1, Secs.~1, 2 and
  Chap.~VII, \S3, Secs.~1, 2 and 4 for a detailed exposition of the
  general method. Elementary function theory then gives an estimate
  for the second derivative of $E(z)$, and Taylor's theorem yields
  \itemref{aptapply}. Details of the argument, geared towards our
  application here, can also be found in Sec.\ 4.1.10 in \cite{Sto01}.

  Concerning observation~\itemref{largedev}, we refer again to analytic
  perturbation theory. The Feynman--Hellmann formula yields 
  \begin{equation}
    E'_{\Lambda}(0)  = \langle \varphi_{0},
    W_{\Lambda}\varphi_{0}\rangle  \,, 
  \end{equation}
  where $\varphi_{0} := |\Lambda|^{-1/2}$, the normalised vector in
  $\ell^{2}(\Lambda)$ with equal components, is the ground state of
  the unperturbed operator $H_{0,\Lambda}$. Therefore, recalling 
  $  W_{\Lambda} = H_{\Lambda} - H_{0,\Lambda} =
  \mathfrak{H}_{\mathcal{G}_{\Lambda}} - \mathfrak{H}_{\LL}$ and
  the definition in \eqref{hlambda}, we have
  \begin{align}
    E^{(\omega)}_{\Lambda}{}'(0) = \frac{1}{|\Lambda|} \;
    \sum_{x\in\Lambda} \sum_{\begin{subarray}{c} y\in\Lambda :\\
        \{x,y\} \in \ELL \setminus \EGL^{(\omega)} \end{subarray}}
    \!\! 1 \;
    &= \, \frac{2}{|\Lambda|} \; \sum_{\{x,y\} \in
      \ELL}   (1- \omega_{\{x,y\}}) \nonumber\\
    & \ge \,\frac{1}{|\ELL|} \; \sum_{\{x,y\} \in \ELL} (1-
    \omega_{\{x,y\}})
  \end{align}
  for all $\omega\in\Omega$.  We recall that the $\omega_{\{x,y\}}$'s, which
  indicate the presence of an edge in the bond-percolation graph, are i.i.d.\ 
  Bernoulli distributed with mean $p$.  Hence, \eqref{ldresult} follows from
  standard large-deviation estimates, see e.g.\ inequality (27.4) in
  \cite{Kal01} or Thm.\ 1.4 in \cite{Tal96}.
 \end{proof}

 So far we have deferred the proof of observation~\itemref{monotone} in the
 above demonstration. This is a
 deterministic result which we address now in

\begin{lem}
  \label{groundstate}
  Let $\Lambda\subset \zd$ be a bounded cube, let
  $\mathfrak{G}=(\Lambda,\mathfrak{E})$ be a subgraph of the fully connected
  cube $\LL$ and let $\HG$ be the finite-volume Laplacian \eqref{hlambda}
  on $\ell^{2}(\Lambda)$. For $t\in\mathbb{R}$ let $\mathfrak{e}(t)$
  be the smallest eigenvalue of 
  \begin{equation}
    \mathfrak{h}(t) := \mathfrak{H}_{\LL} +  t \,\mathfrak{W}\,,
  \end{equation}
  where $\mathfrak{W} := \HG - \mathfrak{H}_{\LL}$. Then
  the function $[0,1] \ni t \mapsto \mathfrak{e}(t)$ is non-decreasing.
\end{lem}

\begin{proof}
  We observe from the definition of $\mathfrak{W}$ and \eqref{hlambda} that
  \begin{equation}
    \label{wrep}
    \mathfrak{W}\,\varphi(x) = \sum_{y \in\Lambda : \{x,y\} \in
      \ELL \setminus {\mathfrak{E}}} \varphi(y)
  \end{equation}
  for all $\varphi\in\ltwoL$ and all $x\in\Lambda$.  Given $t\in [0,1]$ let us
  rewrite $\hate = \HG - (1-t) \,\mathfrak{W} =: 2d\one - \mathfrak{a}(t)$. In
  particular, $\mathfrak{a}(1) = 2d \one - \HG$ acts as
  \begin{equation}
    \label{arep}
    \mathfrak{a}(1)\,\varphi(x) = \sum_{y \in\Lambda : \{x,y\} \in
      \mathfrak{E}} \varphi(y) \; + b_{\partial\Lambda}(x)\varphi(x) 
  \end{equation}
  for all $\varphi\in\ltwoL$ and all $x\in\Lambda$. Equations \eqref{wrep} and
  \eqref{arep} show that the self-adjoint linear operator
  \begin{equation}
    \mathfrak{a}(t) = \mathfrak{a}(1) + (1-t) \,\mathfrak{W}\,,
  \end{equation}
  which lives on the finite-dimensional Hilbert space $\ltwoL$, has only
  non-negative matrix elements $\langle\delta_{x}, \mathfrak{a}(t)
  \delta_{y}\rangle$ for all $x,y\in\Lambda$.  Together with the min-max
  principle, this implies that one can choose the eigenvector(s) corresponding
  to the \emph{largest} eigenvalue of $\mathfrak{a}(t)$ in such a way that all
  their components in the basis $\{\delta_{x}\}_{x\in\Lambda}$ are
  non-negative. Hence, the same is true for the eigenvector(s) corresponding
  to the \emph{smallest} eigenvalue of $\hate$. Thus, another application of
  the min-max principle yields
  \begin{align}
    \mathfrak{e}(t_{2})  &=  \inf_{\substack{0\neq\varphi \in\ltwoL \\
        \varphi(x) 
        \ge 0 \;\;\forall x \in \Lambda}}
    \frac{\langle\varphi,\mathfrak{h}(t_{2})\varphi \rangle}{\langle\varphi,
      \varphi \rangle} 
    = \inf_{\substack{0\neq\varphi \in\ltwoL \\ \varphi(x)
        \ge 0 \;\;\forall x \in \Lambda}}
    \frac{\langle\varphi, \mathfrak{h}(t_{1}) \varphi \rangle  + (t_{2}
      -t_{1}) 
      \langle\varphi,\mathfrak{W}\,\varphi \rangle }{\langle\varphi,
      \varphi \rangle} \nonumber\\
    & \ge \mathfrak{e}(t_{1})
  \end{align}
  for all $0 \le t_{1} \le t_{2} \le 1$, because the scalar product involving
  $\mathfrak{W}$ is non-negative by \eqref{wrep}.
\end{proof}


\section{Proof of Theorem~\ref{hove}}
\label{hoveproof}

In this section we prove the van Hove asymptotics of Theorem~\ref{hove}.
Again, it suffices to consider the lower spectral edge, because of the
symmetries \eqref{bcrel}. That asymptotics follows from

\begin{lem}
  \label{hovelemma}
  Assume $d\in\mathbb{N}\setminus\{1\}$ and $p\in]p_{c},1[$. Then there
  exist constants 
  $\varepsilon_{\mathrm{N}}$, $C_{u}$, $C_{l} \in ]0,\infty[$ such
  that 
  \begin{equation}
    \label{hovebounds}
     C_{l} E^{d/2} \le \NN(E) - \NN(0) \le   C_{u} E^{d/2}
  \end{equation}
  holds for all $E\in ]0, \varepsilon_{\mathrm{N}}[$.
\end{lem}

To prove Lemma~\ref{hovelemma} we separate the contribution of the
percolating cluster to $\NN$ from that of the finite clusters.

\begin{defn}
  Let $\Omega_{\infty}$ denote the event that the origin belongs to the
  percolating cluster and, for $E\in\mathbb{R}$, define 
  \begin{equation}
    \label{NNinf}
    \NNinf (E) := \int_{\Omega_{\infty}}\!\mathbb{P}(\d\omega) \; 
    \langle\delta_{0} ,  \Theta\bigl(E - \DN^{(\omega)}\bigr)
    \delta_{0}\rangle \,, 
  \end{equation}
  which is the contribution of the percolating cluster to the integrated
  density of states of the Neumann Laplacian. We write
  $\widetilde{N}_{\mathrm{N}, \infty}(t) := \int_{0}^{\infty}\!\d
  N_{\mathrm{N}, \infty}(E) \; \e^{-Et}$ for its Laplace transform, where $t
  \in [0,\infty[$.
\end{defn}

As is well known, the Laplace transform of \eqref{NNinf} can be related to the
mean return probability of a continuous-time, simple random walk $\{Z_{t}\}_{t
  \in[0,\infty[}$ on the percolating cluster.  More precisely, this random
walk is the Markov process on $\mathbb{Z}^{d}$ defined by the following set of
rules: Suppose the process is at $x\in\mathbb{Z}^{d}$. Then, after having
waited there for an exponential time of parameter one, one of the $2d$
neighbours of $x$ in $\mathbb{Z}^{d}$, say $y$, is chosen at random with
probability $1/(2d)$. If $\omega_{\{x,y\}}=1$, then the process jumps
immediately to $y$, otherwise there will be no move. The procedure then starts
afresh. Assuming that $Z_{0} =x_{0} \in\mathbb{Z}^{d}$ is the starting point
of the process, we denote its law by $\mathcal{P}_{x_{0}}^{(\omega)}$.  The
process $Z_{t}$ is generated by the Neumann Laplacian in the sense that the
transition probability for going from $x$ to $y$ within time $t$ is given by $
\mathcal{P}_{x_{0}}^{(\omega)}(Z_{s+t}=y \,|\, Z_{s} =x) =
\mathcal{P}_{x}^{(\omega)}(Z_{t}=y) = \langle\delta_{y}, \e^{-t
  \DN^{(\omega)}/(2d)} \delta_{x}\rangle$ for all $s\in [0, \infty[$ and all
$x_{0}$ in the same connected component as $x$ and $y$. From this it follows
that
\begin{equation}
  \label{walk}
  \widetilde{N}_{\mathrm{N}, \infty}(t) = \int_{\Omega_{\infty}}
  \mathbb{P}(\d\omega) \; \mathcal{P}^{(\omega)}_{0}(Z_{2d t} =0)\,.
\end{equation}
Hence, $\bigl(\mathbb{P}(\Omega_{\infty})\bigr)^{-1}\widetilde{N}_{\mathrm{N},
  \infty}(t)$ is the (conditional) mean return probability at time $2dt$ for
the process on the percolating cluster for $p\in ]p_{c},1]$.

Averaged transition probabilities of $Z_{t}$ or related random walks 
have recently been studied in \cite{MaRe04,Bar04,HeHo05} with
elaborate methods. We state a special case of the results as
 
\begin{prop}
  \label{heatkernel}
  Assume $d \in\mathbb{N}\setminus\{1\}$ and $p \in ]p_{c},1[$. Then there
  exist constants $c_{l}, c_{u} \in]0,\infty[$ and $t_{0} \in]1,\infty[$,
  all of which depend only on $p$ and $d$, such that
  \begin{equation}
    \label{heatbounds}
    c_{l} t^{-d/2} \le \widetilde{N}_{\mathrm{N}, \infty}(t)  
    \le c_{u} t^{-d/2}
  \end{equation}
  holds for all $t \in [t_{0},\infty[$.
\end{prop}

\begin{rem}
  In view of \eqref{walk}, the lower bound in
  Proposition~\ref{heatkernel} can be found as Eq.~{\normalfont (30)}
  in Appendix~D in \cite{MaRe04}. That paper also contains a
  ``quenched'' upper bound, i.e.\ an upper bound for
  $\mathcal{P}^{(\omega)}_{x}(Z_{t} =y)$, which is valid for
  $\mathbb{P}$-almost every $\omega\in\Omega_{\infty}$.
  Unfortunately, it is not clear how to take the probabilistic
  expectation thereof, that is, to get an ``annealed'' upper bound. On
  the other hand, the authors of \cite{HeHo05} prove an annealed upper
  bound in their Thm.~{\normalfont 8.1}.  But this bound includes an
  additional logarithmic factor. The strongest results, both annealed
  and quenched, are those of Barlow \cite{Bar04}, and
  \eqref{heatbounds} is a special case thereof. However, his results
  apply to a random walk which is generated by $D^{-1}\DN$ instead of
  $\DN$. Hence, some additional comments are needed to adapt his
  results, and we address this issue now.
\end{rem}

\begin{proof}[Proof of Proposition~\ref{heatkernel}]
  According to the preceding remark the proposition is established, if we show
  that Barlow's quenched upper bound for the return probability, i.e.\ the
  special case $x=y$ of the upper bound in Thm.~2 in \cite{Bar04}, applies
  also to the random walk generated by $\DN$. Eventually, the upper bound in
  Thm.~2 in \cite{Bar04} is reduced to Prop.~3.1 in \cite{Bar04} via
  Thm.~1, Prop.~6.1, Thm.~5.7 and Thm.~3.8 -- the latter being nothing but the
  off-diagonal generalisation of Prop.~3.1, so we do not need it here.  The
  reduction does not make use of any specific properties of the random walk's
  generator.  Hence, all that remains to check is Prop.~3.1. in \cite{Bar04}.
  It turns out that some of the constants in the proof of Prop.~3.1 must be
  modified for our purpose, but this does not have any consequences. In
  addition, the proof of Prop.~3.1 also requires estimates (1.7) and (1.8)
  of Lemma~1.1 in \cite{Bar04}. Estimate (1.8) follows from estimate (1.5), as
  is argued in the proof of Lemma 1.1(b) in \cite{Bar04}. This argument
  applies in our situation, too. So, in the end, we have to verify the validity
  of (1.7) and (1.5) in \cite{Bar04} for the random walk generated by $\DN$.
  
  Estimate (1.5) is seen to hold as a special case of Cor.~11 in \cite{Dav93}.
  The upper bound in estimate (1.7) can be inferred, for example, from
  Thm.~II.5 in \cite{Cou96}, which is a general result for ultracontractive
  Markov semigroups. (Actually, we refer to the first of the two theorems with
  the same number II.5 in \cite{Cou96}.)  To verify statement (ii) of that
  theorem, one may use the application following it together with the
  reasoning in Cor.~V.2 and its proof with the choice $\psi(x) =x$.  This
  choice corresponds to a weak isoperimetric inequality which merely reflects
  that the percolating cluster contains an infinitely long path. From this
  point of view, the weak $t^{-1/2}$-decay of the upper bound in estimate
  (1.7) does not come as a surprise. Finally, the lower bound in estimate
  (1.7) arises solely from the uniform growth condition $|B(x,R)| \le
  \mathrm{const.}  R^{d}$ for the volume of a ball around $x\in\mathbb{Z}^{d}$
  with radius $R$.  Such type of results are well known for heat kernels on
  manifolds and also for discrete-time random walks
  \cite{LuP95,CoGr97,CoGr98}. To employ them here, we decompose
  \begin{equation}
    \mathcal{P}_{x}(Z_{t}=x) = \sum_{n=0}^{\infty} \; \langle\delta_{x},
    K^{n}\delta_{x}\rangle \;\mathcal{P}_{x}\biggl(
    \begin{array}{l}
      \text{there are $n$ attempted}\\[-1.7ex] \text{jumps up to time $t$}
    \end{array}
\biggr)\,,
  \end{equation}
  using the stochastic independence of all building blocks of $Z_{t}$.
  Here, the contraction $K:= \one - \DN/(2d)$ is the transition matrix of a
  discrete-time random walk on $\mathbb{Z}^{d}$, which controls the directions
  of the jumps of $Z_{t}$. The number of attempted jumps up to time $t$ is
  governed by a Poisson distribution with mean $t$. Using this, the lower
  bound in (1.7) follows from Thm. 3(ii) in \cite{LuP95} applied to $K$.
\end{proof}

In order to apply Proposition~\ref{heatkernel} in the proof of
Lemma~\ref{hovelemma}, we will use two elementary Tauberian inequalities.

\begin{lem}
  \label{tauber}
  Let $\mu$ be a positive Borel measure on $\mathbb{R}^{+} =
  [0,\infty[$. Suppose there are constants $t_{0}, \delta, c_{l}, c_{u} \in
  ]0,\infty[$ such that the Laplace transform  
  $\widetilde{\mu}(t) := \int_{\mathbb{R}^{+}}\mu(\d E)\,
  \e^{-Et}$ exists for all $t \in [t_{0},\infty[$ and obeys
  \begin{equation}
    \label{laplacebounds}
    c_{l} t^{-\delta} \le \widetilde{\mu} (t) \le c_{u} t^{-\delta} \,.
  \end{equation}
  Then there exist constants $C_{l}, C_{u} \in ]0,\infty[$ such that 
  \begin{equation}
    \label{measurebounds}
    C_{l} E^{\delta}  \le \mu\bigl([0, E]\bigr) \le C_{u} E^{\delta}
  \end{equation}
  holds for all $E \in ]0, t_{0}^{-1}]$.
\end{lem}

\begin{proof}
  We express $\mu([0, E]) = \int_{\mathbb{R}^{+}}\mu(\d\lambda) \, \Theta (1-
  \lambda /E)$ in terms of the right-continuous Heaviside unit-step function
  and observe the elementary inequality $ \e^{-\tau x} - \e^{-(\tau -1)}
  \e^{-x}\le \Theta(1-x) \le \e^{1-x}$, which is valid for all
  $x\in\mathbb{R}^{+}$ and all $\tau \in [1, \infty[$. The upper bound in
  \eqref{measurebounds} is obvious now. For the lower bound, one has to choose
  $\tau$ large enough such that the constant, which arises from the
  application of both estimates in \eqref{laplacebounds}, is strictly positive.
\end{proof}

\begin{proof}[Proof of Lemma~\ref{hovelemma}]
  We set $\NNfin := \NN -\NNinf$ for the contribution of the finite clusters
  to the integrated density of states and observe
  \begin{equation}
    \label{NNdecomp}
    \NN(E) -\NN(0) = [\NNfin(E) - \NNfin(0)] + \NNinf(E)    
  \end{equation}
  for all $E\in\mathbb{R}$. 
 
  Proposition~\ref{heatkernel} and Lemma~\ref{tauber} establish the desired
  van Hove bounds for $\NNinf$. Therefore, it remains to show that the finite
  clusters do not spoil this behaviour.  Since $\NNfin(E) - \NNfin(0) \ge 0$
  for all $E \in [0, \infty[$, only an appropriate upper bound is required for
  the contribution of the finite clusters. We shall show in Eq.\ 
  \eqref{fincont} below that $\NNfin(E) - \NNfin(0)$ obeys even a
  Lifshits-type upper bound, which will then complete the proof.
  
  The Lifshits behaviour for the contribution of the finite clusters in the
  percolating phase arises from the cluster-size distribution -- in the same
  way as it was shown to arise in the non-percolating phase in \cite{KiMu04}.
  Indeed, we have
  \begin{equation}
    \label{KiMu2.7}
    \NNfin(E) - \NNfin(0) \le \mathbb{P} \bigl\{ \omega \in
    \Omega_{\mathrm{fin}} 
    : |\mathcal{C}_{0}^{(\omega)}| \ge (dE)^{-1/2}\bigr\}
  \end{equation}
  for all $E \in ]0,\infty[$. Here $\Omega_{\mathrm{fin}} := \Omega\setminus
  \Omega_{\infty}$ is the event that the origin belongs to a finite cluster,
  say $\mathcal{C}_{0}^{(\omega)}$, and
  $|\mathcal{C}_{0}^{(\omega)}|$ denotes the number of its vertices.
  Inequality \eqref{KiMu2.7} follows from repeating the steps that lead to the
  first inequality in Eq.~(2.24) in \cite{KiMu04} with $\NNfin$ instead of
  $\NN$. For $p>p_{c}$ the cluster-size distribution on the right-hand side of
  \eqref{KiMu2.7} decays sub-exponentially according to Thm.\ 8.6.5 in
  \cite{Gri99} so that we obtain
  \begin{equation}
    \label{fincont}
    \NNfin(E) - \NNfin(0) \le c_{1} \exp\{ -\xi E^{-(d-1)/2d}\}
  \end{equation}
  for all $E\in ]0,\infty[$ with some constants $c_{1}, \xi\in ]0,\infty[$,
  which depend only on $p$ and $d$. This completes the proof.
 \end{proof}


\end{document}